\documentclass[reprint,aip,jcp,preprintnumbers,superscriptaddress,twocolumn,floatfix]{revtex4-1}

\usepackage[utf8]{inputenc}
\usepackage[T1]{fontenc}
\usepackage{graphicx}
\usepackage{color}
\usepackage{bm}
\usepackage{bbm}
\usepackage{amsmath}
\usepackage{xspace}
\usepackage{bbold}
\usepackage{units}
\usepackage{dcolumn}
\usepackage{paralist}
\usepackage{natmove}
\usepackage{amssymb}
\usepackage{setspace}
\usepackage{mathtools}
\usepackage{gensymb}
\usepackage{siunitx}
\usepackage{booktabs}

\usepackage{tikz}
\usepackage{xcolor}

\begin{document}

\title{Hydration free energies from kernel-based machine learning: Compound-database bias} 
\author{Clemens Rauer}
\affiliation{Max Planck Institute for Polymer Research, 55128 Mainz,
Germany}
\author{Tristan Bereau}
\email{t.bereau@uva.nl}
\affiliation{Max Planck Institute for Polymer Research, 55128 Mainz,
Germany} 
\affiliation{Van 't Hoff Institute for Molecular Sciences and
Informatics Institute, University of Amsterdam, Amsterdam 1098 XH, The
Netherlands}

\date{\today}
\begin{abstract}
  We consider the prediction of a basic thermodynamic
  property---hydration free energies---across a large subset of the
  chemical space of small organic molecules. Our in silico study is
  based on computer simulations at the atomistic level with implicit
  solvent. We report on a kernel-based machine learning approach that
  is inspired by recent work in learning electronic properties, but
  differs in key aspects: The representation is averaged over several
  conformers to account for the statistical ensemble. We also include
  an atomic-decomposition ansatz, which we show offers significant
  added transferability compared to molecular learning. Finally, we
  explore the existence of severe biases from databases of
  experimental compounds. By performing a combination of
  dimensionality reduction and cross-learning models, we show that the
  rate of learning depends significantly on the breadth and variety of
  the training dataset. Our study highlights the dangers of fitting
  machine-learning models to databases of narrow chemical range.
\end{abstract}

\maketitle

\section{Introduction}


Applications of machine-learning (ML) models to atomic and molecular
systems have had tremendous impact in our ability to tackle a more
systematic exploration of chemical compound
space~\cite{behler2016perspective, ramakrishnan2017machine,
sanchez2018inverse, von2018quantum}. Much of these developments stem
from a combination of apt representations that incorporate the
relevant symmetries together with flexible and expressive
interpolation machines~\cite{bartok2013representing,
schutt2017quantum, huo2017unified, chmiela2017machine,
glielmo2017accurate, grisafi2018symmetry, bereau2018non,
thomas2018tensor}. Work in the last few years has been devoted to the
learning of electronic properties of molecules: atomization
energies~\cite{rupp2012fast, hansen2013assessment}, multipole
moments~\cite{bereau2015transferable, gastegger2017machine}, the
electron density~\cite{grisafi2018transferable}, or the wave
function~\cite{schutt2019unifying}.

In contrast, applications of machine learning have not transpired as
much to biomolecular systems and soft matter, where configurational
averages can lead to significant entropic
effects~\cite{bereau2016research, ferguson2017machine, Bereau2018,
jackson2019recent}. Recent examples include developments in
coarse-grained force fields~\cite{john2017many, wang2019machine,
chan2019machine, moradzadeh2019transfer}, optimizing collective
variables~\cite{sultan2018automated, wehmeyer2018time,
varolgunes2020interpretable}, as well as compound screening and
optimization~\cite{lee2016mapping, hoffmann2019controlled}. 

Predicting thermodynamic properties across chemical space is of high
industrial and technological relevance. Strong interests in drug
design, for instance predictions of water-octanol partitioning or
protein-ligand binding, are illustrated by decades-old
contributions~\cite{hopfinger1976application, marshall1987computer}.
Computationally efficient predictions of protein-ligand binding
traditionally entail the statistical scoring of a docked ligand in a
protein pocket~\cite{leach2006prediction}. Virtual drug discovery
adopted early on a framework to correlate molecular structure with
physicochemical as well as biochemical
properties~\cite{kubinyi19933d}. More recent applications have
leveraged the use of modern machine-learning techniques to improve
predictive capabilities~\cite{ghasemi2018neural}. The field remains
plagued by limited and noisy reference experimental data, despite
efforts at improving transferability~\cite{altae2017low}. This overall
hinders ML models in reaching satisfying generalization across
chemical space.

At the crux of ML generalization is the breadth and variety of the
chemical space spanned by the training set~\cite{Lin2018,
glavatskikh2019dataset}. Chemical space is overwhelmingly
large---supposedly up to $10^{60}$ drug-like molecules---making any
exhaustive treatment unconceivable~\cite{dobson2004chemical,
reymond2015chemical}. Any compound dataset, typically in the range
$10^3-10^7$ molecules, thereby stands as a minuscule subsampling of
chemical space. How uniform---or at least representative---can such a
subsampling be? Experimental datasets suffer from biases due to both
practical interests in specific interactions (e.g., hydrogen bonds),
as well as historic developments in synthetic
chemistry~\cite{menichetti2019drug}. Recent successes in ML
applications for electronic properties, on the other hand, have
largely stemmed from \emph{dense} subsets of chemical space,
incorporating a rich, representative coverage over a small
neighborhood. Databases such as the GDB algorithmically enumerate
molecules that ought to be chemically
stable~\cite{ruddigkeit2012enumeration}, unlike experimentally
available compounds that are scarcely populated in chemical space.

In this work we focus on a basic yet fundamental thermodynamic
property: hydration free energies (HFEs). HFEs quantify the free
energy required to transfer a solute molecule from vacuum to
bulk-liquid water. We point out the existence of several recent deep
ML models for hydration free energies, including AIMNet based on a
density-based solvation model~\cite{zubatyuk2019accurate},
DeepChem\cite{Hutchinson2019Solvent}, which works on functional class
fingerprints, and Delfos targeting different solutes and
solvents~\cite{lim2019delfos}. The present report consists of a
comparative study of the performance of kernel-based ML models against
three databases:
\begin{itemize}
\item QM9 is based on the algorithmically-grown GDB
  database~\cite{ramakrishnan2014quantum}.  QM9 consists of 134k
  molecules with up to 9 heavy atoms, including chemical elements C,
  O, N, and F. For this study we removed all molecules containing
  fluorine. We restrict our study to 4000 randomly-selected compounds.
\item eMolecules consists of more than 20 million commercially
  available compounds~\cite{emolecules}. We limit the set to up to 9
  heavy atoms and elements C, O, and N. From the resulting 34\,517
  molecules we randomly selected 4000.
\item FreeSolv consisting of around 500 molecules with experimentally
  available HFEs~\cite{mobley2014freesolv}. Further limiting this set
  to up to 9 heavy elements C, O, and N reduces to 259 compounds.
\end{itemize}

Various methods exist to predict HFEs in silico, the main workhorse
being molecular dynamics (MD) together with physics-based force
fields, coupled with rigorous free-energy calculation techniques
(e.g., alchemical transformations)~\cite{shivakumar2010prediction}.
Though explicit-solvent MD simulations remain the best compromise in
terms of accuracy and transferability, they remain computationally
expensive, preventing us from easily generating large databases.
Furthermore, setting up a protocol for accurate explicit-solvent
simulations requires extreme care, at odds with a screening
study~\cite{gaieb2018d3r}. Instead we turn to \emph{implicit}-solvent
MD simulations to generate our reference free energies.
Implicit-solvent simulations run in the gas phase and add a
Poisson-Boltzmann solvation term to the
Hamiltonian~\cite{genheden2015mm}. They display larger errors compared
to explicit-solvent simulations (2.6 versus
1.3~kcal/mol~\cite{nicholls2008predicting}), but at a significantly
lower computational cost. 

To enhance the generalization of the ML model, we explore two aspects.
First, rather than feeding the representation of a single conformer,
our representation averages over \emph{several} snapshots---a proxy
for the underlying configurational average. Physically any arbitrary
configuration is devoid of any statistical weight, it is only the
configurational average that ought to link to the free energy. Second,
we probe the ability to learn \emph{atomic} contributions of the free
energy via an additive decomposition ansatz. Despite the absence of
physical justification, we propose it in an effort to reduce the
underlying interpolation space, and will empirically test its ability
to improve transferability.

We first describe the theoretical setting, in particular the
kernel-based ML modeling. We describe the formalism behind the
atomic-decomposition ansatz and test its transferability.
Atom-decomposed ML models will be compared to simple linear regression
as baseline to better grasp the requirements on the training set.
Finally, we compare the learning performance in the three databases,
operate cross-learning between databases to probe their
transferability, and study the breadth and variety of the spanned
chemical space through dimensionality reduction.

\section{Methods}


\subsection{Linear Model}
\label{sec:lin}

To later assess the quality of our ML models, we first propose the use
of linear regression as a baseline. We express the molecular free
energy of hydration (HFE), $G$, as a linear combination of atomic
contributions, weighted by the number of corresponding atoms in the
compound. For molecule $i$ this would correspond to
\begin{equation}
  G_i = \sum_j N_i^j g^j,
  \label{eq:lin}
\end{equation}
where $N_i^j$ is the number of atoms of type $j$ in molecule $i$, and
$g^j$ is the contribution of atom type $j$ to the molecular free
energy. We can then write the linear system as a matrix equation ${\bm
G} = N{\bm g}$, where $N$ is the matrix of atomic contributions and
${\bm g}$ is the unknown vector of atomic contributions.

We consider two models with different numbers of atomic parameters:
($i$) four chemical elements (C, O, N, H) and ($ii$) 39 atom types of
the GAFF force field, as assigned by the force-field generating {\sc
Antechamber} program~\cite{wang2001antechamber}.

\subsection{Kernel-ridge regression}

We use kernel-ridge regression (KRR) to learn the mapping ${\bm Q}
\mapsto G$, where ${\bm Q}$ denotes an input molecular representation
and $G$ its corresponding HFE. The two quantities can be linked via a
kernel, $\hat K_{ij} = \hat K({\bm Q}_i,{\bm Q}_j) = {\rm Cov}({\bm
G}_i,{\bm G}_j)$, that encodes the similarity between inputs ${\bm
Q}_i$ and ${\bm Q}_j$. Training a kernel model is equivalent to
solving the set of linear equations ${\bm G} = \hat K {\bm \alpha}$,
where ${\bm \alpha}$ is the vector of weight coefficients. This vector
is optimized by inversion of the Tikhonov-regularized problem
\begin{equation}
  {\bm \alpha} = (\hat K + \lambda \mathbbm{1})^{-1} {\bm G},
  \label{eq:krr}
\end{equation}
with hyperparameter $\lambda$ and the identity matrix $\mathbbm{1}$.
Prediction for compound ${\bm Q}^*$ is subsequently obtained through
an expansion of the kernel evaluated on the training set
\begin{equation}
  G({\bm Q}^*) = \sum_{i=1}^N \alpha_i \hat K({\bm Q}_i, {\bm Q}^*),
\end{equation}
where index $i$ runs over all $N$ training points.

We use a Gaussian kernel with Euclidean norm
\begin{equation}
K({\bm Q}_i, {\bm Q}_j) = 
\exp\left( -\frac{||{\bm Q}_i-{\bm Q}_j||_2^2}
{2\sigma^{2}}\right),
 \label{eq:kernel}
\end{equation}
where hyperparameter $\sigma$ defines the width of the kernel
distribution. The molecular representation, ${\bm Q}$, used in this
work is the Spectrum of London and Axilrod-Teller-Muto (ATM) potential
(SLATM).\cite{huang2017dna, Huang2018} It encodes each atom $i$ of a
molecule via a histogram of 1-, 2-, and 3-body terms in the
neighborhood of atom $i$, stored in one vector. Each level of
interaction encodes respectively ($i$) the nuclear charge $Z_i$,
($ii$) the spectrum of radial distribution of the London potential
$\rho^{ij}_i(r)$, and ($iii$) the spectrum of angular distribution of
the ATM potential $\rho^{ijk}_i(\theta)$, where $r$ is a distance,
$\theta$ is an angle, and $i \neq j \neq k$. The representation is
invariant to translation, rotation, and permutation. A molecular
representation is thereby built through the concatenation of each atom
vector. Both molecular and atomic variants of this representation
exist, and are used here. SLATM describes a single molecular
configuration, while free energies inherently result from a
configurational average. Therefore, we adapt the representation used
to be not just the SLATM vector of one conformation, but be the
\emph{average} of the SLATM vectors over 30 Boltzmann-weighted
snapshots sampled every 100ps from the gas phase simulations used to
calculate the HFEs (see Section \ref{section_implicit}. Working with
the average, rather than a concatenation, allows us to bypass any
ordering issue of the conformations in the representation vector. 

\subsection{Atomic-decomposition ansatz}
\label{sec:met_ansatz}

The above-mentioned KRR scheme aims at the prediction of HFEs for an
entire molecule at once---one pair of molecules per entry in the
kernel matrix $\hat K$. As an alternative, we explore the possibility
to learn \emph{atomic} contributions to the HFE. We generalize our
linear models (Sec.~\ref{sec:lin}) such that each atomic contribution
may not be strictly limited in resolution to a chemical element alone,
but more broadly its local environment, which we refer to here as an
\emph{atom-in-molecule} contribution. This approach will benefit from
a smaller interpolation space, thereby facilitating learning
\cite{Ceriotti2018, Scherer2020}. Expressing the HFE of a molecule
from atomic contributions de facto assumes a decomposition
\begin{equation}
  \label{eq:atomic_dec}
  G({\bm Q}) = \sum_{l=1}^{n_{\rm atoms}} g({\bm q}_l),
\end{equation}
where ${\bm q}_l$ is the atom-in-molecule representation of atom $l$
and $g({\bm q}_l)$ is its contribution to the molecular HFE, $G$. We
will refer to Eq.~\ref{eq:atomic_dec} as the atomic-decomposition
ansatz. Effectively the aSLATM representation of ${\bm q}_l$
generalizes the concept of atom types in Eq.~\ref{eq:lin}.

We aim at establishing a second mapping ${\bm q} \mapsto g$ via a
\emph{local} kernel $\hat k$. Eq.~\ref{eq:atomic_dec} links the two
kernels, $\hat K$ and $\hat k$. The target properties available for
the global kernel will allow us to infer an ML model for the local
kernel, despite the lack of atomic target properties. We rewrite
Eq.~\ref{eq:atomic_dec} as a set of $N$ molecules with a corresponding
set of $M$ atomic contributions by introducing the mapping matrix
$\hat L$
\begin{equation}
  {\bm G} = \hat L {\bm g}.
\end{equation}
The coefficient $\hat L_{ij}$ is 1 if molecule $i$ contains atomic
contribution $j$, and 0 otherwise. In case that molecule $i$ contains
$n$ identical atomic contributions $j$, this would lead to a
coefficient $\hat L_{ij} = n$, reducing the size of the matrix. This
bookkeeping matrix allows us to link the global and local kernels
$\hat K = \hat L \hat k \hat L^\intercal$~\cite{Scherer2020}. Training
an ML model takes advantage of both this relationship between kernels
and the linear system of equation ${\bm G} = \hat K {\bm \alpha}$
\begin{equation}
  {\bm \alpha} = \left( \hat L \hat k \hat L^\intercal 
  + \lambda \mathbbm{1}\right)^{-1} {\bm G}.
\end{equation}
Once trained, predictions of both atomic contributions and molecular
free energies are respectively given by
\begin{align}
  {\bm g}^* &= \left( \hat L \hat k^* \right)^\intercal {\bm \alpha} \\
  {\bm G}^* &= \hat L^* \left( \hat L \hat k^* \right)^\intercal 
  {\bm \alpha},
\end{align}
where $\hat k^*$ and $\hat K^*$ refer to the local and global kernels
between training and test data, respectively.

In the following, $\hat k$ takes the same form as $\hat K$ (see
Eq.~\ref{eq:kernel}). For the atomic representation we use the atomic
variant of SLATM, denoted herein aSLATM.

The hyperparameters that need optimization consist of the
regularization term $\lambda$ in Eq.~\ref{eq:krr} as well as the
length-scale normalization $\sigma$ of the Gaussian kernel
(Eq.~\ref{eq:kernel}). A systematic grid search led to the values
$\sigma = 100$ and $\lambda = 10^{-8}$. We found the latter being very
much insensitive to the accuracy of the kernel over a wide range of
values. For all KRRs the results shown are averaged over 5 independent
runs using random train-test splits.

\subsection{Computer simulations}\label{section_implicit}

Computer simulations were carried out in {\sc Gromacs}
2016.4~\cite{Abraham2015}. We generated initial molecular
configurations using \texttt{rdkit} starting from their SMILES
string~\cite{rdkit}. The force field was generated from the {\sc
Antechamber} program package with AM1-BCC atomic
charges~\cite{wang2001antechamber, da2012acpype}. We ran gas-phase
molecular dynamics simulations using Langevin dynamics at $T=300$~K
for 3~ns, of which we omit the first \SI{100}{ps} for equilibration.
We average over 29 evenly-spaced snapshots of the trajectory, and
compute the HFE from the molecular mechanics Poisson-Boltzmann surface
area (MM-PBSA), as implemented in
\texttt{g\_mm/pbsa}\cite{Kumari2014gmmpbsa, Baker2001Electrostatics}.
The polar contribution was obtained using the Poisson-Boltzmann solver
with the vacuum, solute and solvent dielectric constants set to
$\varepsilon = 1$, $2$, and $80$, respectively. The nonpolar solvation
energy was calculated using the solvent accessible surface area method
using a probe radius $r = 1.4$~\AA$^3$, surface-tension parameter
$\gamma = 0.0226778$~kJ mol$^{-1}$ \AA$^{-2}$, and an offset $\Delta
G_{\rm corr} = 3.84982$~kJ mol$^{-1}$.


\section{Results}

\subsection{Reference free energies}

\begin{figure}[htbp]
 \includegraphics[width=.9\linewidth]{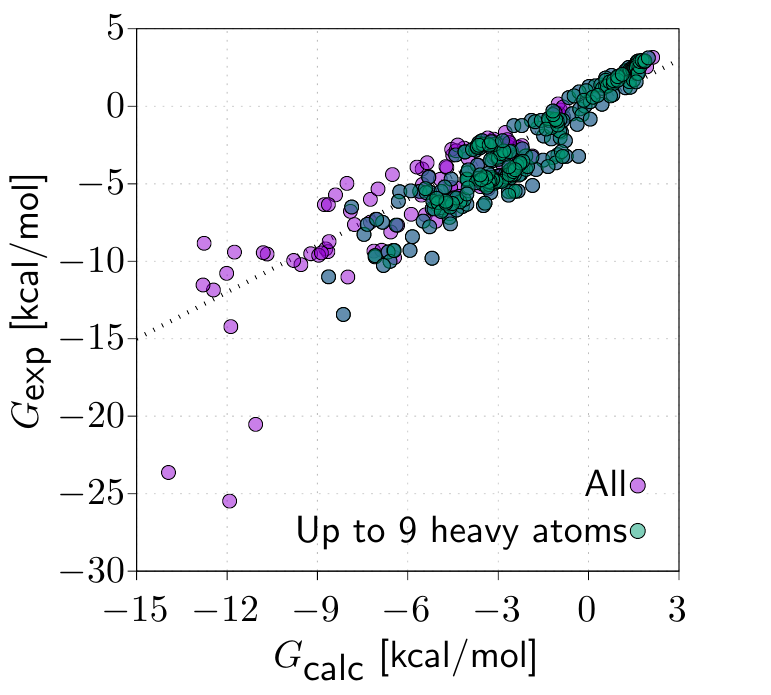}
 \caption{Correlation of experimental, $G_{\rm exp}$, and
 implicit-solvent, $G_{\rm calc}$, hydration free energies (HFEs) for
 355 CHON molecules from the FreeSolv database. The subset of points
 in green focuses on compounds up to 9 heavy atoms. The dotted line
 indicates perfect agreement.}
 \label{fig_impl_exp}
\end{figure}

We first benchmark our implicit-solvent calculations against
experimental HFEs. We focus on a subset of 355 molecules from the
FreeSolv database, consisting solely of chemical elements C, H, O, and
N (herein denoted CHON). Fig.~\ref{fig_impl_exp} shows a correlation
of the HFEs from both simulations and experiments. The mean absolute
error (MAE) of the implicit solvation HFEs of these molecules is $1.29
\pm 1.24$ kcal/mol. The standard deviation is heavily impacted by
outliers of large molecular weight. However, the databases we work
with in this study contain mostly compounds up to 9 heavy atoms, which
feature a lower standard deviation (highlighted in
Fig.~\ref{fig_impl_exp})---MAE of $1.24 \pm 0.86$ kcal/mol.

\begin{figure}[htbp]
 \includegraphics[width=.9\linewidth]{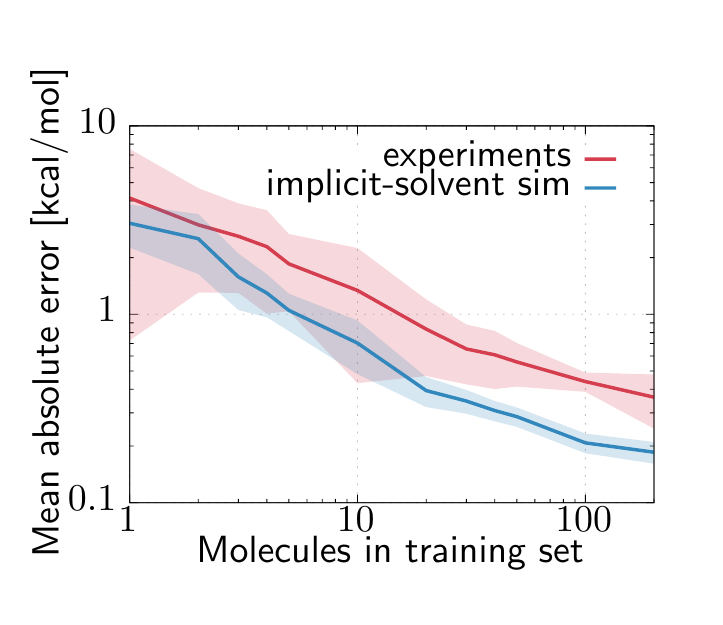}
 \caption{Learning curves for the HFE of the freeSolv dataset using both the experimental and the calculated implicit-solvent free energies. Both curves are averaged over 5 independent simulations.}
 \label{fig_impl_exp_lc}
\end{figure}

We further compare the consistency of calculated implicit-solvent and
experimental free-energy datasets by comparing learning curves of ML
models. KRR was applied to learn reference free energies obtained by
each data set. We use the FreeSolv dataset with up to 9 heavy atoms.
For both learning procedures, we rely on the same averaged aSLATM
vectors obtained from the conformational ensemble of the simulations.
The results for both sets of HFEs are shown in
Fig.~\ref{fig_impl_exp_lc}. The implicit-solvation free energies yield
better learning performance than the experimental values. Despite a
virtually identical slope, the experimental predictions are shifted
\emph{up} by roughly $0.3$ kcal/mol. We point at two possible reasons:
($i$) This shift can be affected by the conformational sampling being
identical for both curves, leading to more consistency for the
calculated implicit-solvent free energies, as conformational sampling
and free energies were obtained from the same set of simulations.
($ii$) However, it could also be caused by the experimental free
energies being more heterogeneous. All in all, the results show that
implicit-solvent calculations offer a reasonable proxy for the
experimental values, which we rely on in the rest of this work.

In the Supporting Information we provide CSV files containing all
reference free energies calculated as basis of this work.

\subsection{Atomic-decomposition ansatz}

\begin{figure}[htbp]
	\includegraphics[width=.9\linewidth]{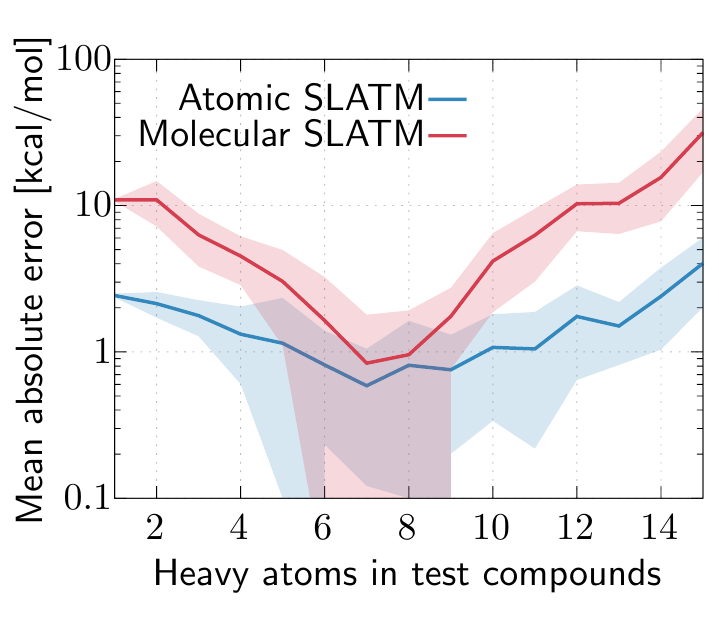}
    \caption{Atomic and Molecular SLATM representations using a KRR
    model learned on 1000 QM9 compounds with 7 and 8 heavy atoms.
    Predictions on the FreeSolv database.}
    \label{fig_mol_atom}
\end{figure}

We set out to compare the learning performance of HFEs from the
molecular representation and the atomic-decomposition ansatz exposed
in Section~\ref{sec:met_ansatz}. As a challenging test we specifically
focus on learning both types of transferability simultaneously: across
databases and compound size. We first picked 663 molecules out of QM9,
with 86 and 577 compounds featuring 7 and 8 heavy atoms, respectively.
Using this training set we optimized both a global and a local ML
model to predict implicit-solvent HFEs. We apply the two ML models to
predict the implicit-solvent HFEs of a different set of 351 molecules
taken out of the FreeSolv database, which feature a broad range of
molecular weights. Fig.~\ref{fig_mol_atom} displays the mean absolute
error of the two ML models as a function of the number of heavy atoms
in the predicted (test) compounds. Overall, we find that the global ML
model $\hat K$ leads to significantly larger errors---up to a factor
of 5 in error compared to the atomic kernel for the smallest and
largest compounds. The atomic-decomposed ML model, on the other hand,
features a slight improvement around 7 and 8 heavy atoms---used for
the training set. Furthermore, it displays a remarkably flat behavior
(shown on a logarithmic scale), indicating robust transferability
across molecular weight. We do observe an increase of the error for
molecules toward 15 heavy atoms, likely hinting at a lack of coverage
of sufficient chemical environments in the training. The effect may be
compounded with significant errors of the implicit-solvent method for
large compounds, suggesting a lack of coherence across molecular
weight.

The results empirically justify the atomic-decomposition ansatz: the
ML model using the aSLATM atom-in-molecule representation reaches
better performance across molecular weights, despite the lack of
formal justification for Eq.~\ref{eq:atomic_dec}. Indeed, the free
energy is an ensemble property of the \emph{entire} system. 
A decomposition of $G$ into finer components may be considered, for
instance via the thermodynamic integration (TI)
formula.\cite{Kirkwood1935} TI couples the two Hamiltonians of the
solute without and with solvent
\begin{equation}
  G = \int_0^1 {\rm d}\lambda \left\langle 
  \frac{\partial U}{\partial \lambda} \right\rangle_\lambda,
\end{equation}
where $U$ is the potential energy of the system, $\lambda$ is a
coupling parameter between the two Hamiltonians, and
$\langle\cdot\rangle_\lambda$ denotes an average over the canonical
ensemble at coupling parameter $\lambda$. For most explicit-solvent
atomistic force fields, the non-bonded interactions of $U$ are
pairwise decomposable. In our case, the use of MM-PBSA clouds a simple
decomposition, due to the solvation term. However, establishing links
between TI and an atom-decomposed ML model may help to shed light on
key contributions of the free energy.

The present results indicate that any error due to the decomposition
ansatz must be smaller than the accuracy of learning reached from this
small dataset. The behavior of the atomic-decomposed ML model as a
function of training-set size is probed next.

\subsection{Atomic decomposition: Rate of learning}

In order to test the performance of our atomic-decomposed ML model, we
take as a training set 4000 random molecules sampled from the QM9
database for which we have calculated the HFE using MM/PBSA.

As a baseline model we predict the HFEs using linear regression, with
both the 4 elements CHON and the 39 GAFF atom types. Both models were
trained on the HFEs of 2000 randomly-selected molecules. We report the
mean absolute errors for the held-out 2000 compounds in
Tab.~\ref{tab_linear}. The 4-element CHON model has an MAE of $1.80
\pm 1.98$ kcal/mol, while the refined 39-atom-type GAFF model yields
$1.06 \pm 0.02$ kcal/mol. As such, splitting chemical environments in
the 39 atom types defined by GAFF almost yield chemical accuracy
(1~kcal/mol).

\begin{table}[htbp]
 \begin{tabular}{cccc}
 \toprule
   Model name & Parameters & MAE (kcal$/$mol) \\
   \midrule
   CHON elements & 4 & 1.80 $\pm$ 1.98 \\
   GAFF atom types & 39 & 1.06 $\pm$ 0.02 \\
   \bottomrule
 \end{tabular}
 \caption{Mean absolute error for both CHON and GAFF linear regression
 models. Both models were trained and tested on a 50\% hold-out split
 of 4000 QM9 molecules.}
 \label{tab_linear}
\end{table}

\begin{figure}[htbp]
 \includegraphics[width=0.45\textwidth]{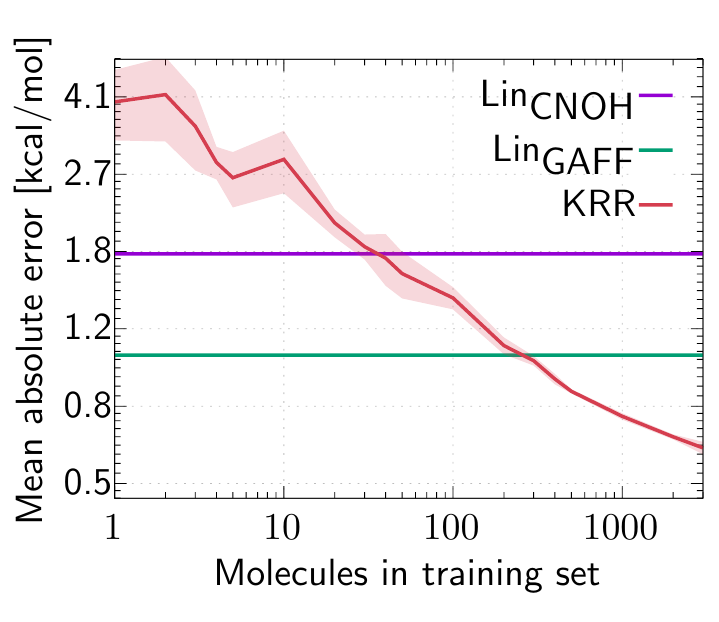}
 \caption{Learning curve of the hydration free energies for 4000
 randomly chosen molecules of the QM9 database. The two lines show
 the prediction error for the CHON and GAFF linear models. KRR is the
 atomic-decomposed ML model.}
 \label{fig_gaff}
\end{figure}

These linear models offer us a way to better assess the performance of
the atomic-decomposed ML model. Fig.~\ref{fig_gaff} compares the
different regressions in terms of their out-of-sample predictions,
displaying the MAE as a function of the training-set size. The
atom-decomposed ML model needs 40 and 300 molecules to outperform the
CHON and GAFF linear models, respectively. With 2500 molecules in the
training set, the atom-decomposed ML model yields an MAE of only
0.7~kcal/mol. It illustrates how offering a richer description
consistently improves the performance: from chemical elements (CHON),
to a select list of force-field-based atom types (GAFF), to the
atom-in-molecule aSLATM representation. The latter can be seen as a
continuous generalization of the others, offering a more accurate
mapping between local chemical environment and free-energy
contribution. 

%

\subsection{Database bias}
\label{section_databases}

If learning HFEs across a subset of QM9 requires $\sim 10^2$ training
compounds, how transferable is this result to other databases? We set
the stage for this question by comparing atom-decomposed learning in
three different databases, which differently try to span chemical
space: QM9, eMolecules, and FreeSolv. We will more directly address
the question further down in Sec.~\ref{sec:crosslearn}.

\begin{figure}[htbp]
 \includegraphics[width=.45\textwidth]
 {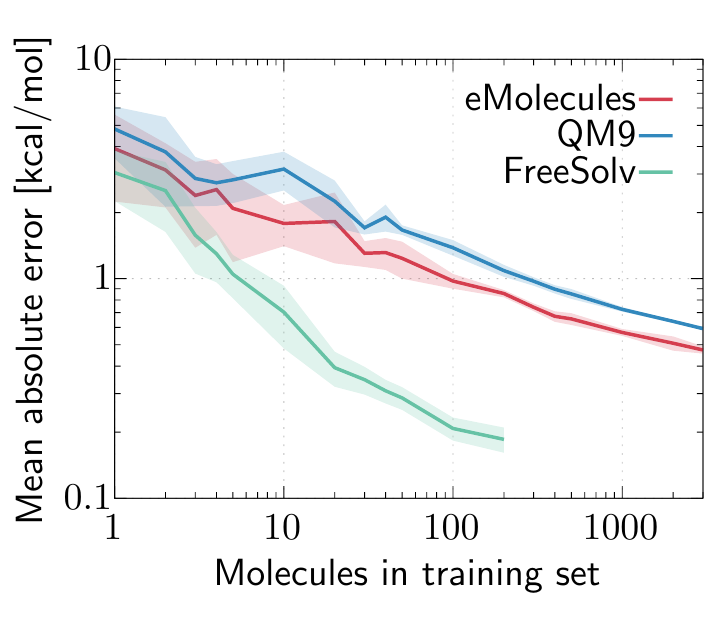} \caption{Learning
 curves for the HFE ML models for the three databases QM9, eMolecules
 and FreeSolv. All curves are averaged over 5 independent runs.}
 \label{fig_databases}
\end{figure}

Figure \ref{fig_databases} shows independent learning curves for the
three databases. While QM9 and eMolecules show similar learning
performance---the latter being slightly more performant---we observe
surprisingly different behavior for FreeSolv: the learning curve
reaches chemical accuracy after less than 10 compounds, and only
0.3~kcal/mol after 200 training molecules. The three ML models
featuring identical architectures and representations, the results
suggest a significant bias in the nature of the subsets of chemical
space they span.

To probe the difference in the spanned chemistries of the three
databases, we performed dimensionality reduction. We used the
\texttt{UMAP} algorithm.\cite{mcinnes2018umap} UMAP builds a fuzzy
topological representation in the original high-dimensional space, and
identifies a low-dimensional embedding by means of a cross-entropy
measure. The UMAP parameters consisted of the number of neighbors
(15), the minimum distance (0.1), the dimensionality of the embedding
(1 or 2), and the metric (Euclidean).

\begin{figure}[htbp]
 \includegraphics[width=.45\textwidth]{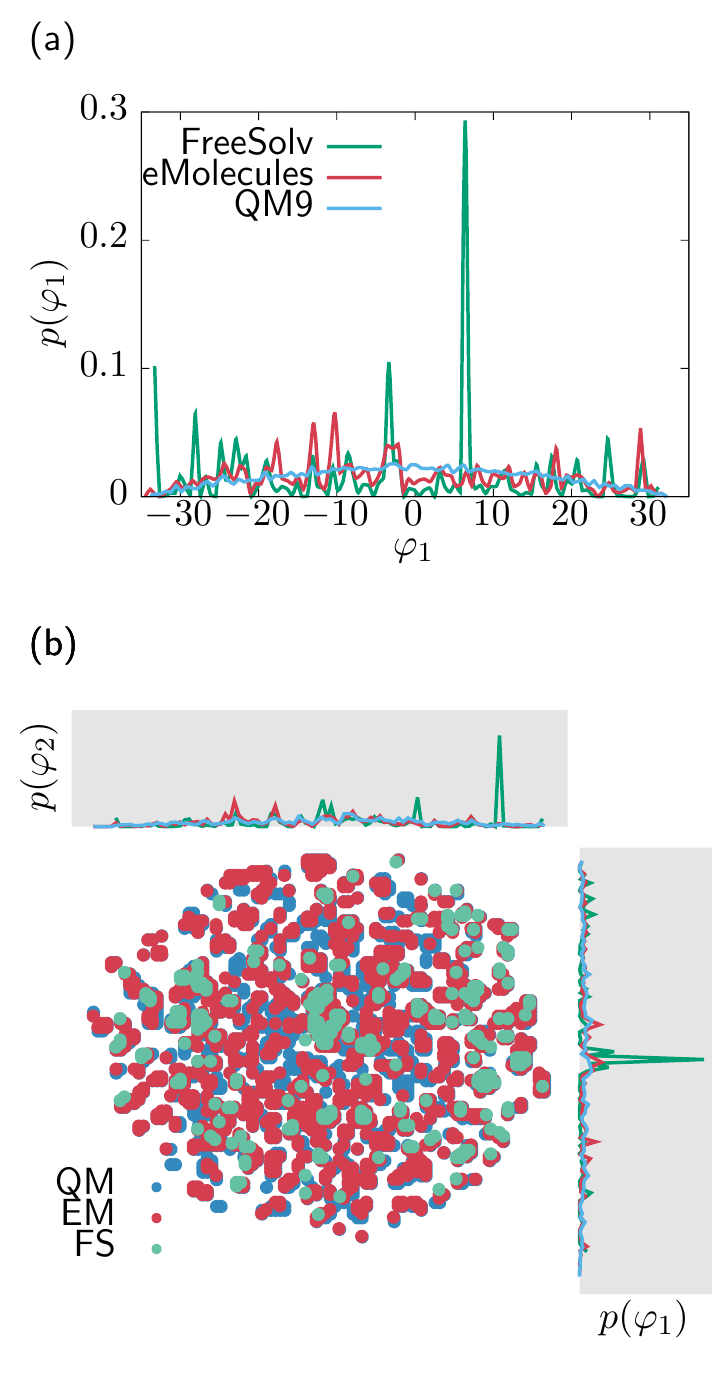}
 \caption{(a) Probability distribution, $p(\varphi_1)$, of the 1D UMAP
 projection of the three databases. (b) Probability distribution,
 $p(\varphi_1, \varphi_2)$, and 1-dimensional cuts of the 2D UMAP
 projection of the three databases QM9 (QM), eMolecules (EM), and
 FreeSolv (FS).}
 \label{fig_hist}
\end{figure}

To compare the three databases we need to project them onto the same
reduced subspace. A one-dimensional (1D) projection was first trained
from the atomic SLATM representation on QM9, as we assumed QM9 to
cover chemical space more broadly than the other two, since it
represents a subset of the GDB. After training we map all three
databases onto this 1D projection, subsequently called $\varphi_1$.
Fig.~\ref{fig_hist}a shows the probability distribution,
$p(\varphi_1)$, as a measure of coverage in that subspace. The results
are striking: While QM9 shows a remarkably flat distribution across
the range of $\varphi_1$, eMolecules displays larger fluctuations,
while FreeSolv yields high peaks, indicating significant localization
within the subspace. This localization translates into significant
bias in the chemical space spanned by the database---most atomic
environments cluster at few points within the $\varphi_1$ subspace.
This behavior explains the exceptional learning behavior shown in
Fig.~\ref{fig_databases}. The intermediate regime of eMolecules
(between broad and spiked) indicates slight but noticeable
localizations in chemical space, translating in learning performance
that is slightly more favorable than QM9. As such the commercially
available database shows less diversity than the algorithmically
generated database.

The same trend applies when mapping to a 2-dimensional (2D) mapping of
the chemical space spanned by these databases. A similar training
procedure is applied as for the 1D case. Fig.~\ref{fig_hist}b shows
the 2D probability distribution, $p(\varphi_1, \varphi_2)$, as well as
1D cuts thereof. The sharp peaks of FreeSolv subsist in \emph{both}
dimensions. The 2D space more explicitly illustrates the presence of
``islands'' in chemical space---most strongly pronounced for
FreeSolv---but we also find significant differences between QM9 and
eMolecules.

\subsection{Cross-Learning}
\label{sec:crosslearn}

\begin{figure*}[htbp]
\includegraphics[width=0.95\linewidth]{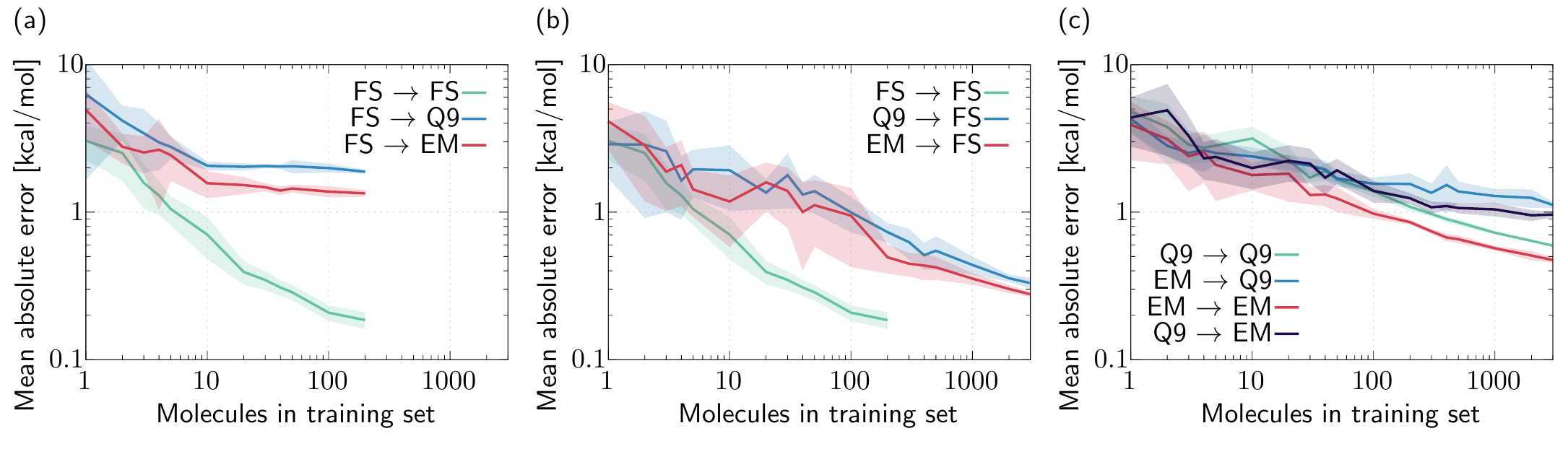}
\caption{Learning curves for the cross-learning experiments: training
 on one dataset and prediction on another. FS, EM, and QM stand for
 FreeSolv, eMolecules, and QM9, respectively. The legends indicate the
 type of cross-learning: $X \to Y$ means training on $X$ and
 prediction on $Y$.}
\label{fig_cross_learn}
\end{figure*}

To further probe the overlap between subsets of chemical space spanned
by the three datasets, we performed cross-learning experiments, in
which we train on one dataset and predict on another. For FreeSolv all
259 molecules were used for the training and test sets, while for QM9
and eMolecules we considered all 4000 molecules for training, and up
to 3000 randomly-selected compounds for testing.
Fig.~\ref{fig_cross_learn} and Tab.~\ref{tab_cross_learn} report the
cross-learning results for our ML models, which we analyze in the
following.

\begin{table}[htbp]
 \begin{tabular}{cccc}
 \toprule
 Training & \multicolumn{3}{c}{Prediction} \\
 \cmidrule{2-4}
   & QM9 & eMolecules & FreeSolv \\
 \midrule
  QM9 & $0.65\pm 0.02$ & $1.00\pm 0.03$ & $0.36\pm 0.02$  \\
  eMolecules & $1.26\pm 0.05$ & $0.61\pm 0.01$ & $0.35\pm 0.01$ \\
  FreeSolv & $2.04\pm 0.06$& $1.26\pm 0.05$ & $0.24\pm 0.02$ \\
  \bottomrule
 \end{tabular}
\caption{Cross-learning machine learning models of the three databases
considered in this study. All energies given in kcal/mol.}
\label{tab_cross_learn}
\end{table}

Fig.~\ref{fig_cross_learn}a reports cross-learning trained on the
FreeSolv database. The FreeSolv $\to$ FreeSolv is identical to what
was shown in Fig.~\ref{fig_databases}, and cross-learning to the other
databases shows a significant deterioration of the accuracy: QM9 and
eMolecules saturate to roughly 1.9 and 1.3~kcal/mol. We recall the
results from Fig.~\ref{fig_gaff}: the CHON and GAFF linear models
reach an accuracy of 1.80 and 1.06~kcal/mol. As such the FreeSolv
cross-learning on QM9 is \emph{worse} than a 4-parameter linear
regression. Beyond the MAEs at highest training size, what is striking
is the apparent plateau behavior after the first decade of training
points: Learning improves negligibly from 10 to $10^2$ data points.
This aspect speaks for the lack of breadth of chemical space---the
dataset features few, overrepresented chemical environments.
Furthermore, the small but noticeable offset between eMolecules and
QM9 suggests more difficulties in learning the latter, which further
hints at its broader diversity of chemical environments.

Panel b of Fig.~\ref{fig_cross_learn} demonstrates the opposite
effect: to what extent the three databases can predict FreeSolv. All
three databases eventually lead to similar accuracy, albeit with
different learning rates: FreeSolv is more efficient at learning
itself than the others, while QM9 and eMolecules show significant
offsets. This is another hint at the broader diversity of compounds
from eMolecules and, to a larger extent, QM9. 

Finally, Fig.~\ref{fig_cross_learn}c focuses on the comparison between
eMolecules and QM9. All curves roughly start at the same offset at low
training data. Both self-learning curves (QM9 $\to$ QM9 and eMolecules
$\to$ eMolecules) reach the lowest MAE, thanks to a slightly more
favorable rate of learning. When it comes to cross-learning, QM9 shows
a slight advantage at learning eMolecules than vice versa. Based on
the dimensionality reduction (Fig.~\ref{fig_hist}), we argue that this
arises from the broader diversity of chemical environments present in
QM9.

\section{Conclusions}

In the present work we study the machine learning (ML) of hydration
free energies (HFEs) across a subset of small organic
molecules---those made of chemical elements C, H, O, and N. To probe
the effects of database biases on the learning of thermodynamic
properties, we generated reference HFEs using implicit-solvent
computer simulations at an atomistic resolution. We find that an
atomic-decomposition ansatz, in which we assume a linear decomposition
of the HFE in atomic contributions, offers remarkable transferability,
compared to the more straightforward learning of the molecular
property. As baseline we compare two linear models based on atom
types, as often used in force fields. A 39-parameter model based on
the GAFF atom types yields 1.06~kcal/mol. Training a better performing
atom-decomposed ML model requires a couple hundred molecules in the
training set. The atom-in-molecule environment encoded in the aSLATM
representation offers de facto a generalization of the concept of
force-field atom types.

ML models trained on different databases show significantly different
performance. Using dimensionality reduction, we find that FreeSolv and
eMolecules, two databases of commercially available compounds, show
strong localizations in the chemical space spanned. We can very
efficiently train an ML model out of the FreeSolv database, but it
fails to generalize to the other databases. Furthermore,
cross-learning across databases shows that training a model on
FreeSolv and deploying it on QM9 is worse than a 4-parameter linear
model, and shows a severe plateau behavior, highlighting the lack of
chemical diversity. The combination of cross-learning and
dimensionality reduction shows that supervised learning can help
empirically establish which database probes a broader chemical space.
It also shows that deploying an ML model on independent databases can
help probe its generalization.

\section{Supplementary Material}

CSV data files for subsets of the FreeSolv,\cite{mobley2014freesolv}
eMolecules,\cite{emolecules} and QM9\cite{ramakrishnan2014quantum}
databases containing SMILES strings and associated hydration free
energies, as calculated from atomistic simulations with implicit
solvent. The data file for FreeSolv additionally contains reference
experimental free energies.\cite{mobley2014freesolv}

\section{Acknowledgements}
We thank Bernadette Mohr and Roberto Menichetti for critical reading
of the manuscript. We also thank Denis Andrienko, Kiran H.~Kanekal,
and Christoph Scherer for insightful discussions. This work was
partially supported by the Emmy Noether program of the Deutsche
Forchungsgemeinschaft (DFG). TB acknowledges the Institute for Pure
and Applied Mathematics (IPAM), which is supported by the National
Science Foundation (Grant No. DMS-1440415).

\section{Data availability}

The data that supports the findings of this study are available within
the article and its supplementary material.


\bibliographystyle{unsrt}

\bibliography{biblio}


\end{document}